\documentclass{article}
\usepackage[latin9]{inputenc}
\usepackage{float}
\usepackage{multirow}
\usepackage{amsmath,bm,amssymb}
\usepackage{graphicx}
\usepackage{hyperref}
\usepackage{ragged2e}
\usepackage{microtype}
\usepackage{subfigure}
\usepackage{booktabs} % for professional tables
\usepackage{tikz}
\usetikzlibrary{positioning}
\justifying

\makeatletter

\providecommand{\tabularnewline}{\\}

\usepackage{spconf}
\usepackage{epstopdf}
\usepackage{epsfig}
\usepackage{multicol}
\usepackage{comment}
\usepackage{url}

\title{EAT: Enhanced ASR-TTS for Self-supervised Speech Recognition}
\name{\parbox{0.9\linewidth}{\center 
Murali Karthick Baskar$^\phi$, Luk\'{a}\v{s} Burget$^\phi$, Shinji Watanabe$^\dagger$, Ramon Fernandez Astudillo$^\pi$,  \newline and Jan ``Honza'' \v{C}ernock\'{y}$^\phi$}  \thanks{
 All the authors from Brno university of Technology are supported by Czech National Science Foundation (GACR) project "NEUREM3" No. 19-26934X and Czech Ministry of Education, Youth and Sports project No. LTAIN19087 "Multi-linguality in speech technologies".
}}
\address{
$^\phi$Brno University of Technology,  $^\dagger$ Johns Hopkins University, $^\pi${IBM Research}, \\
  {\small \tt {baskar@fit.vutbr.cz}}
}

\makeatother

\begin{document}
\ninept \maketitle 
\begin{abstract}
Self-supervised ASR-TTS models suffer in out-of-domain data conditions. Here we propose an enhanced ASR-TTS (EAT) model that incorporates two main features: 1) The ASR$\rightarrow$TTS direction is equipped with a language model reward to penalize the ASR hypotheses before forwarding it to TTS. 2) In the TTS$\rightarrow$ASR direction, a hyper-parameter is introduced to scale the attention context from synthesized speech before sending it to ASR to handle out-of-domain data. Training strategies and the effectiveness of the EAT model are explored under out-of-domain data conditions. The results show that EAT reduces the performance gap between supervised and self-supervised training significantly by absolute 2.6\%  and 2.7\% on Librispeech and BABEL respectively.
\end{abstract}

\begin{keywords} cycle-consistency, self-supervision, sequence-to-sequence, speech recognition \end{keywords}

\section{Introduction}
The application of the sequence-to-sequence architecture~\cite{bahdanau2014neural} to ASR and TTS models paved way to perform self-supervised training by simple integration of ASR and TTS. Recent works on self-supervised training~\cite{tjandra2017listening,hori2018cycle,baskar2019semi,hayashi2018back} leveraging unpaired speech and text have shown higher performance compared to other unsupervised training approaches. Most of the research in self-supervised ASR is done in effectively integrating ASR and TTS such that it is differentiable~\cite{tjandra2018end} and easily trainable. However, ASR and TTS are exploited in disconnected fashion by synthesizing speech using TTS~\cite{rossenbach2020generating,sun2020generating,du2020speaker} and improving ASR through data augmentation. These techniques focus on the synthesis part and rely on text only data from unpaired sets to improve recognition performance. The  work in~\cite{Kahn_2020} also improves ASR performance, by using a language model as a hypothesis scorer and applying self-training techniques over the resulting corrected pseudo-labels. In ~\cite{baevski2019effectiveness}, the authors apply self-supervision through pre-training with the help of a BERT model to improve ASR performance with unpaired data. BERT has also been used as a effective pre-training technique with contrastive loss in~\cite{baevski2020wav2vec} by training in self-supervised fashion.

A recent work~\cite{masumura2020sequence} on semi-supervised sequence-to-sequence ASR has applied consistency training and has shown effectiveness with unlabeled speech data. Our previous work called ASR-TTS~\cite{baskar2019semi} used cycle-consistency training with REINFORCE and showed gains on standard speech datasets. However, our experiments with other corpora showed that the model suffers under the out-of-domain data condition and has further room for improvement in-terms of training and architecture. 

In this work, we investigate methods to improve the robustness of the cycle-consistency approach in limited data and out-of-domain scenarios. The contributions can be itemized as follows

\begin{itemize}
    \item We incorporate a pre-trained RNNLM regularization term in the ASR REINFORCE loss for speech only (SO) training, increasing its robustness to bad latent ASR hypotheses.
    \item We introduce a hyper-parameter for text-only (TO) training, to attenuate the influence of the ASR encoder by scaling the attention-encoded context. This allows us to reduce the focus on acoustic information when the latent speech quality is poor, effectively alternating between ASR and a more language-model-like behaviour.
    \item We incorporate latest training strategies and architectures such as data augmentation and data annealing. The TTS module is also built using the Transformer architecture, which shows higher robustness and memory efficiency. Multi-head attention is used as ASR encoder layers to attain additional gains and for reduced model complexities.
    \item We show that these techniques greatly improve performance and particularly attain the target goal achieving good performance in limited data and out-of-domain scenarios with cycle consistency techniques. 
\end{itemize}

%related work here: write about tjandra, fb where only librispeech is used. and how our work is different with babel data

We call this improved model \emph{enhanced ASR-TTS (EAT)}. Experiments are conducted on Librispeech and the BABEL-Pashto datasets and show the efficacy of our EAT model. Results are further compared with state-of-the-art (SotA) results in literature.

\section{Preliminaries}
Our previous work, ASR-TTS~\cite{baskar2019semi} is a self-supervised training system built to handle both speech only (SO)
and text only (TO) data using a cycle-consistency training regime. ASR-TTS training approach containing two pipelines: 1) ASR$\rightarrow$TTS pipeline to train using SO dataset 2) TTS$\rightarrow$ASR pipeline to train using TO dataset. The SO data $x$ is fed to ASR$\rightarrow$TTS pipeline to reconstruct $x$ as $\hat{x}$. The pipeline is trained with an expected loss (approximated through REINFORCE). The text only data $y$ is fed to the TTS$\rightarrow$ASR pipeline to predict the text $\hat{y}$ and is trained with a cross-entropy loss objective. Both  pipelines act as  auto-encoders allowing to perform self-supervised training on either SO dataset $\mathcal{D}_{u}^{s}$ or TO dataset $\mathcal{D}_{u}^{t}$. The ASR architecture used in our ASR-TTS model is built using an RNN based sequence-to-sequence model and the Tacotron is used for TTS.

\section{Enhanced ASR-TTS (EAT)}
The EAT proposed in this work includes two main modifications to our previous work~\cite{baskar2019semi}, first on SO and other on TO training in the ASR-TTS model. 
\begin{figure}[h]
\begin{center}
\includegraphics[scale=0.5]{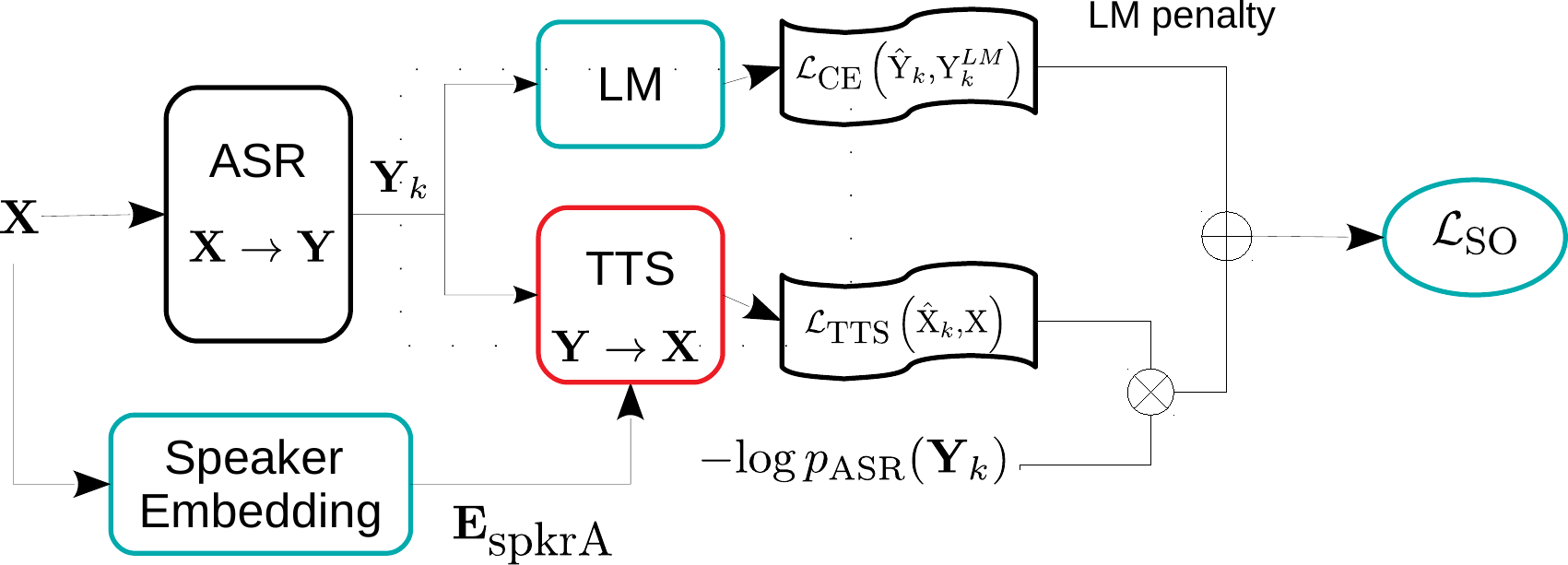}
%\label{fig:asr2tts}
\caption{Speech Only (SO) data training using ASR $\rightarrow$ TTS pipeline with LM penalty.}\label{fig:asr2tts}
\end{center}
\end{figure}
\vspace{-0.2cm}
\subsection{Adding a RNNLM penalty for regularization}
The ASR$\rightarrow$TTS cycle-consistent SO training objective used  in~\cite{baskar2019semi}
\begin{equation}
    \mathcal{L}_{\mathrm{SO}}^{'}= E_{p_{\mathrm{ASR}}(\mathbf{Y}\mid \mathbf{X})}\{\mathcal{L}_{\mathrm{TTS}}(\mathbf{X}\mid \mathbf{Y})\}\label{eq:1},
\end{equation}
is the expected TTS negative log-likelihood $\mathcal{L}_{\mathrm{TTS}}(\mathbf{X}\mid\mathbf{Y})$ for the latent ASR hypotheses $\mathbf{Y}$. Note that this likelihood is teacher-forced i.e. the ground truth is used for the auto-regressive component. One limitation of this approach is that cycle-consistency may not be restrictive enough to avoid erroneous hypotheses for $\mathbf{Y}$, making training less robust. To solve this, we incorporate a $\beta$-weighted negative log likelihood of a RNN language model to equation~\eqref{eq:1} as shown in figure~\ref{fig:asr2tts} and yielding 

\begin{equation}
    \mathcal{L}_{\mathrm{SO}} = E_{p_{\mathrm{ASR}}(\mathbf{Y}\mid \mathbf{X})}\{\mathcal{L}_{\mathrm{TTS}}(\mathbf{X}\mid \mathbf{Y}) + \beta \mathcal{L}_{\mathrm{LM}}(\mathbf{Y})\}\label{eq:2},
\end{equation}
which plays a regularization role similar to the Kullback-Leibler term in Variational Auto-encoders (VAEs)~\cite{kingma2013auto}\footnotemark\footnotetext{Note the similarity with the Evidence Lower Bound (ELBO) $\log p(x) \geq E_{q(y \mid x)}\{\log p(x \mid y) + \log p(y) - \log q(y \mid x)\}$}. The expectation is approximated with REINFORCE~\cite{williams1992simple}.

\subsection{Making TTS\texorpdfstring{$\rightarrow$}ASR robust to out-of-domain}

The TTS$\rightarrow$ASR cycle-consistent TO training objective from~\cite{baskar2019semi} exhibits a major weakness when training with out-of-domain data. TTS is less robust to out-of-domain data and generates poor log-Mel filterbank (fbank) frames in this condition. In this pipeline, features $\hat{\mathbf{X}}$ are predicted by TTS as
\begin{equation}
    \hat{\mathbf{X}} = \arg\max_{\mathbf{X}}\{p_{\mathrm{TTS}}(\mathbf{X}\mid \mathbf{Y})\}\label{eq:4}
\end{equation}
encoded in the ASR encoder as $H=\mbox{Encoder}(X)$, and sent to the attention component to obtain the attention context vector $c_{l}$ as

\begin{equation}
c_{l}=\sum_t{a_{lt}h_{t}}\label{eq:5}
\end{equation}
 where $t$ and $l$ denote the time step and token id respectively. The final loss is then given by 
\begin{equation}
\mathcal{L}_{\mathrm{TO}}= -\mbox{log}\,p_{\mathrm{ASR}}(\mathbf{Y}^{*}\mid \hat{\mathbf{X}})=-\sum_{l}^{L}\log\mbox{Decoder}(c_{l},y_{l-1})\label{eq:6}
\end{equation}

\begin{figure}[ht]
\begin{center}
 \begin{tikzpicture}
  \node (img)  {\includegraphics[scale=0.9]{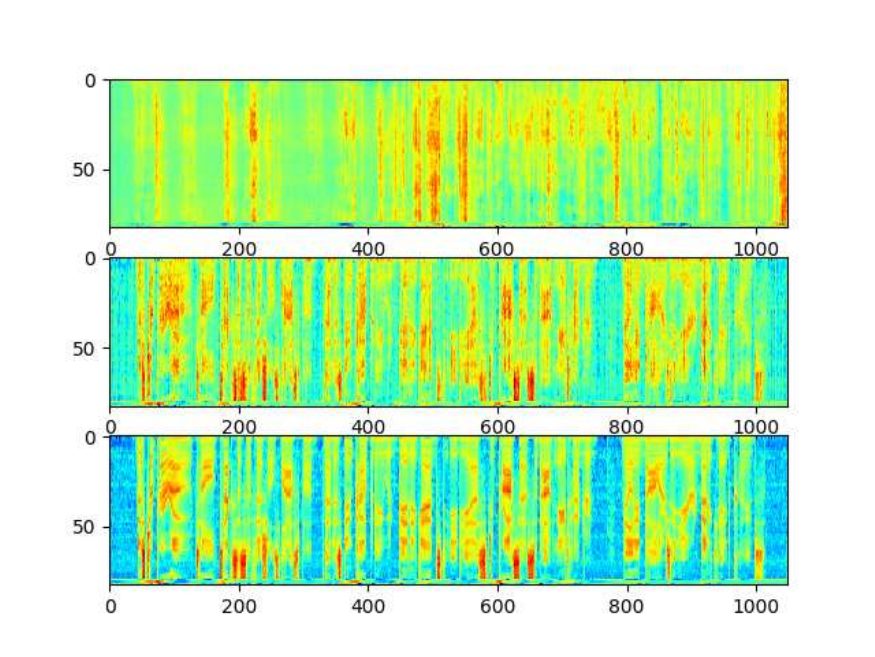}};
 \node[below=of img, node distance=0cm, yshift=1cm,font=\color{black}] {\# frames};
  \node[left=of img, node distance=0cm, rotate=90, anchor=center,yshift=-0.7cm,font=\color{black}] {\# log-Mel filter-banks};
 \end{tikzpicture}
\caption{Plot of log-Mel filter-bank features. The top plot shows the features predicted by TTS during TTS$\rightarrow$ASR training. The plot in the middle shows the features predicted by TTS during ASR$\rightarrow$TTS training. The bottom figure shows the ground-truth of the log-Mel filterbank (fbank) features.}
\label{fig:spec}
\end{center}
\end{figure}
The top plot in figure~\ref{fig:spec} shows the reconstructed fbank features of Librispeech using a TTS pre-trained with WSJ data.  Comparing the top and middle plots, corresponding TTS$\rightarrow$ASR and ASR$\rightarrow$TTS training respectively, one can see a clear difference in prediction error.

The primary reason behind this is that the ground truth is available in ASR$\rightarrow$TTS to perform teacher-forcing. Whereas in TTS$\rightarrow$ASR, this is not available for TTS and thus the reconstructed output deviates from the ground truth. Even the segments of speech and silence are wrongly predicted in the TTS$\rightarrow$ASR pipeline.
\begin{figure}[H]
\begin{center}
\includegraphics[scale=0.6]{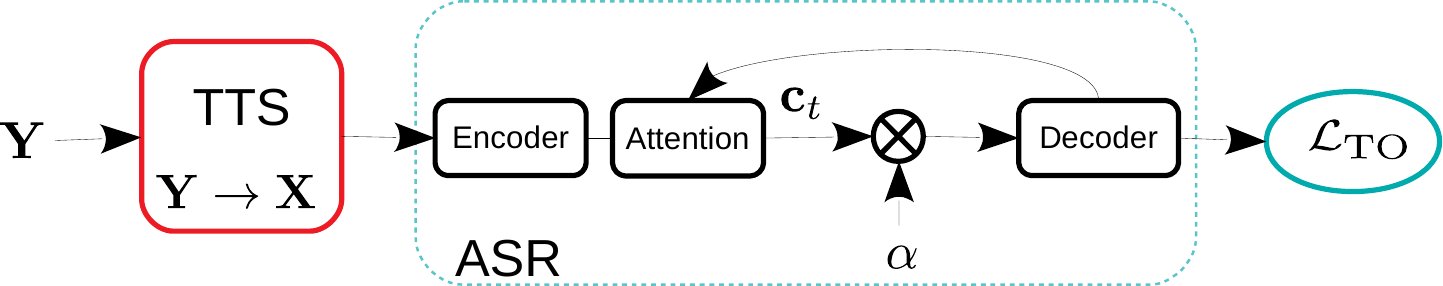}
\caption{Text only (TO) data training with inclusion of attention context scaling by $\alpha$ in TTS$\rightarrow$ASR pipeline}\label{fig:tts2asr}
\end{center}
\end{figure}
\vspace{-0.3cm}

In EAT, we mitigate this robustness issue by scaling the attention context vector in the ASR module by a hyper-parameter $\alpha$ 

\begin{equation}
c_{l}=\alpha \sum_t{a_{lt}h_{t}}\label{eq:7}
\end{equation}

also shown in figure~\ref{fig:tts2asr}. If $\alpha=0$, no encoder features are used and the ASR model just behaves as a language model. This prevents the erroneous TTS generated features to provide a misleading signal, while still allowing to backpropagate into the ASR decoder. The value of $\alpha$ is chosen heuristically based on the difference in domains between data used to train TTS and the TO data. The final loss, used both for speech only and text only data (ST), is given by summing the loss functions $\mathcal{L}_{\mathrm{ST}} = \mathcal{L}_{\mathrm{SO}} + \mathcal{L}_{\mathrm{TO}}$  of the above pipelines.

\subsection{Improvements in architecture and training}
\subsubsection{Model Architecture}
The ASR and TTS architecture in the EAT model is meticulously designed as it plays a major role in attaining improved performance. The motivation is to keep the model light-weight and simple to easily fit in GPU memory during training.

\textit{ASR}:
The ASR component in EAT model is equipped with location based multi-head attention component~\cite{karita2019comparative}. Instead of RNN layers in encoder, self-attention layers are used to reduce model complexity. The decoder is built with RNN layers as before, since the transformer decoder is harder to implement with our training objective. In our experiments, we use 2 VGG layers followed by 6 self-attention layers each with 800 dimensions. The encoder output is sent to attention component with 10 heads and 512 dimensions. 10 convolution channels with 100 filters are used in this attention to be location specific. Adadelta is used and trained with batch size 20. Our experiments shows that multi-head attention and self-attention layer based encoder provided performance gains.

\textit{TTS}:
Transformer based TTS~\cite{hayashi2020espnet,li2019neural} is used in this work, as we found in our experiments that Transformer consumes less memory and is effective in out-of-domain condition when compared to Tacotron architecture~\cite{hayashi2020espnet}. The TTS is multi-speaker based and handles each speaker input by providing an x-vector~\cite{snyder2018x} as speaker embedding. The Transformer architecture contains 6 encoder and decoder layers each with 1536 units respectively. The attention component contains 4 attention heads, each with 384 attention dimension. 2 pre-net layers with 256 units and 5 post-net layers with 256 channels are used. The output frames reduction factor is set to 1 as all frames are required during self-supervised training. Speaker embeddings are added to the encoder output before sending to the decoder. The pre-training of transformer TTS is done using its standard optimizer with 10000 warmup steps for 100 epochs.

\subsubsection{Data Augmentation}
In ASR-TTS, simple Gaussian noise is used as augmentation to stabilise the training and it provided minor gains but was inconsistent across datasets. In this work, the EAT model is trained with the specaugment~\cite{park2019specaugment} approach. The frequency mask and time mask are applied using a window width of 30 consecutive log-Mel frequency channels and 40 consecutive time steps respectively. The recognition performance of the EAT model with specaugment is shown in table~\ref{tab:aug}. Specaugment approach attains consistent gains using self-supervised training with SO and ST (SO + TO) data. No augmentation is done during training with TO data.

\begin{figure}[h]
\texttt{in \textbf{general} investors are a conservative \textbf{lot} these days she says} (ground-truth)\\
\texttt{in \textbf{\textcolor{blue}{general}} investors are a conservative \textbf{\textcolor{red}{lat}} these days she says} (baseline) \\
\texttt{in \textbf{\textcolor{red}{deneral}} investors are a conservative \textbf{\textcolor{blue}{lot}} these days she say} (ASR-TTS)
\caption{An example of text sequence predicted by baseline and ASR-TTS compared with the ground-truth}
\label{fig:seq}
\end{figure}
\vspace{-0.5cm}
\subsubsection{Data Annealing}

In ASR-TTS model training, alternating between large amounts of unsupervised data and little supervised data is difficult. The supervised training of certain labels can result in over-fitting, which hinders the effect of unsupervised training~\cite{xie2019unsupervised} as shown in figure~\ref{fig:seq}.

Here, "general" in reference text is correctly predicted in baseline training. During ASR-TTS training, the supervised samples from baseline and unsupervised samples are provided in alternate fashion. Although, in~\cite{baskar2019semi} we repeated the supervised data to reduce the under-fitting, it still resulted in incorrect predictions such as "deneral" and also increased the training time due to repetition. To mitigate this, the supervised samples are released only when:
\vspace{-0.2cm}
%\begin{equation}
\begin{align}
p_{\mathrm{ASR}}(\hat{y}\mid x) > {\gamma}_t; {\gamma}_t = {\eta}_t\times(1-\frac{1}{K}) + \frac{1}{K} \times {\eta}_t\\
%%$K$ is the number of classes.
{\eta}_t \rightarrow 1-\exp(\frac{t}{T}\ast5),\,\, \text{  log schedule} \\
{\eta}_t \rightarrow \exp((\frac{t}{T}-1)\times5),\,\, \text{  exp schedule}
\end{align}
%\vspace{-0.2cm}
%\end{equation}
where $T$ and $K$ are the number of training steps and classes respectively. Table~\ref{tab:sch} shows that linear and exp schedules are better over log, as release of supervision is initially high and reduces at the end of training. The performance of EAT trained using exp schedule outperforms linear as the supervised data is mostly released at the final stage of training paving smoother way for training with unsupervision.

\section{Results and discussion}

Librispeech~\cite{panayotov2015librispeech} and BABEL-Pashto~\cite{karafiat2016multilingual} datasets are used in our experiments. The WSJ-si84 is used to pre-train ASR and TTS models. 83 dimensional filterbank features are extracted and used to train our ASR and TTS systems. EAT model training is performed under SO, TO and ST condition by splitting the data into unpaired and paired data. 100 hours of Librispeech is used as paired data and 360 hours of Librispeech as unpaired data. 5 and 10 hours of paired data are obtained from 39.74 hours of BABEL-Pashto data and the rest of the dataset is used as unpaired data as denoted in table~\ref{tab:pas}. Baseline models denoted in table \ref{tab:aug} and \ref{tab:pas} are built only with paired data. RNNLM for Librispeech is built with 460 hours of clean and 500 hours of other data. RNNLM for Pashto is built with  external text containing 83k utterances and 62k vocabulary size. Our experiments are done using ESPnet toolkit and the code will be published on github~\footnote{https://github.com/creatorscan/espnet-asrtts}. All experiments are conducted with RNNLM during testing. Evaluation with Librispeech is done on dev-clean, dev-other, test-clean and test-other as such variability can showcase the effectiveness of EAT.

\subsection{Results on Librispeech}
EAT is  initially tested with different data annealing schedules to chose the best training schedule for the rest of the experiments. 360 hours of both speech only and text only (ST) data is used during EAT training and the results for log, linear and exp schedules are in table~\ref{tab:sch}. The results show that for Librispeech, exp schedule is better and will be used in the rest of our experiments.
\vspace{-0.4cm}
\begin{table}[h]
\caption{\%WER performance of log, linear and exp based annealing schedules data during self-supervised training using EAT}
\label{tab:sch}
\begin{tabular}{ccccc}
\hline 
360-ST & dev-clean & dev-other & test-clean & test-other\tabularnewline
\hline 
\hline 
log & 7.7 & 23.5 & 6.9 & 24.3\tabularnewline
linear & 7.1 & 22.7 & 6.9 & 23.6\tabularnewline
exp & 6.9 & 22.5 & 6.9 & 22.1\tabularnewline
\hline 
\end{tabular}
\end{table}

Table~\ref{tab:aug} shows the effect of data augmentation by specaugment. SO training with 360 hours of data attains consistent gains on all evaluation sets. 360-ST denotes that the 360 hours of speech only and text only data are used simultaneously for training EAT. The performance improvements obtained by 360-SO and 360-ST with augmentation shows that the EAT model training is complementary to specaugment approach. Here oracle in the table denotes the performance of ASR model trained with 460 hours of data in Librispeech. The effect of SO and ST data training with RNNLM penalizer (LMP) in EAT is also shown in table~\ref{tab:aug}. The penalty loss aids the model training to reduce incorrect ASR predictions and the results show significant improvement in performance. 360-SO + LMP attains 19.0\%WER on harder evaluation set such as test-other when compared to 25.0\%WER with specaugment only. 
\vspace{-0.5cm}
\begin{table}[h]
\caption{Recognition performance of EAT model using specaugment approach, RNNLM penalizer (LMP) and attention context scaler ($\alpha$)}
\label{tab:aug}

\begin{tabular}{lcccc}
\hline 
Type & dev-clean & dev-other & test-clean & test-other\tabularnewline
\hline 
\hline
Baseline & 14.3 & 36.4 & 14.4 & 36.9\tabularnewline
\hline
360-SO & 11.0 & 32.4 & 10.6 & 33.6\tabularnewline
$\,\,\,$+aug & 9.1 & 24.2 & 8.9 & 25.0\tabularnewline
$\,\,\,$+LMP & 6.0 & 18.6 & 5.8 & 19.0\tabularnewline
\hline 
360-TO & 8.9 & 23.0 & 8.6 & 24.1\tabularnewline
$\,\,\,$+$\alpha$ &4.5 & 15.8 & 4.7 & 15.9\tabularnewline
\hline 
360-ST & 6.9 & 22.5 & 6.9 & 23.6\tabularnewline
$\,\,\,$+aug & 6.0 & 18.6 & 5.8 & 19.0\tabularnewline
$\,\,\,$+$\alpha$ & 5.2 & 19.5 & 5.3 & 20.4\tabularnewline
$\,\,\,$+LMP & 4.3 & 14.9 & 4.3 & 15.2\tabularnewline
\hline 
oracle &3.7 &12.3 &3.5 &12.6\tabularnewline
\hline
\end{tabular}
\end{table}
\vspace{-0.2cm}
The attention context vector in ASR is scaled by $\alpha=0.7$ for Librispeech and attains 15.9\%WER on test-other with 360-TO when compared to 24.1\%WER by 360-TO without $\alpha$ scalar. This simple, yet effective method has allowed 360-ST to further improve its performance which shows that LMP and $\alpha$ scaler are complementary.
In 360-ST, inclusion of $\alpha$ and LMP results in 4.3\% WER on test-clean and 14.9\% WER on test-other. The oracle experiment is done by training an ASR with 460 hours of supervised data and it attains 3.5\%WER on test-clean and 12.6\%WER on test-other.
\vspace{-0.3cm}
\begin{table}[h]
\caption{\%WER of EAT on BABEL-Pashto using aug, $\alpha$ for TO and LMP for SO with 5 and 10 hours of paired data}
\label{tab:pas}
\centering{}%
\begin{tabular}{lcccc}
\hline 
Supervision info & Baseline & SO & TO & ST\tabularnewline
\hline 
5 hours & 71.4 & 63.9 & 62.2 & 63.1\tabularnewline
$\,\,\,$+ aug & 63.1 & 61.4 & 61.7 & 60.1\tabularnewline
$\,\,\,$+ $\alpha$ & - & - & 58.8 & 58.5\tabularnewline
$\,\,\,$+ LMP & - & 60.1 & - & 58.4\tabularnewline
\hline
10 hours & 64.7 & 61.1 & 60.9 & 60.4\tabularnewline
$\,\,\,$+ aug & 55.5 & 55.0 & 54.0 & 53.6\tabularnewline
$\,\,\,$+ $\alpha$ & - & - & 53.7 & 52.5\tabularnewline
$\,\,\,$+ LMP & - & 54.8 & - & 51.6\tabularnewline
\hline
Oracle & 56.0 & - & - & -\tabularnewline
$\,\,\,$+ aug & 48.9 & - & - & - \tabularnewline
\hline
\end{tabular}
\end{table}
\vspace{-0.6cm}
\subsection{Results on BABEL-Pashto}
The key results of the EAT are shown on BABEL-Pashto as it helps to show the impact of $\alpha$ and LMP when compared to our previous ASR-TTS model. Experiments on Pashto using our previous model did not lead to gains and hence are not included in this paper. The reason behind the difficulty is that building a multi-speaker TTS model for Pashto is harder and hence our previous work failed to provide reasonable TTS scores. EAT model mitigates this problem by modifying its TTS architecture and reducing the importance of synthesized speech from TTS by $\alpha$. Here, the pre-trained TTS is retrained during SO with RNNLM penalizer and later used for TO training with $\alpha=0.3$.
Table~\ref{tab:pas} shows that with 5 hours of paired data, the effect of TO is higher compared to SO, but with 10 hours the TO obtains comparable gains as SO. With $\alpha$ in TO the model obtains 58.8 \%WER and further reduced to 58.5 \%WER with ST training. Here oracle in table denotes the performance of ASR trained with 39.75 hours of data in Pashto. With 10 hours of paired data and ST training with both $\alpha$ and LMP, the EAT model attains 51.6 \%WER which is only absolute 2.7\% less compared to oracle 48.9 \%WER.
\vspace{-0.4cm}
\subsection{Comparison with related work}
Some of the recent works using ASR and TTS to handle unpaired speech and text data have raised the performance bar on Librispeech. The self-training approach such as pseudo-label training~\cite{synnaeve2019end} feeds the SO data to a pre-trained ASR and uses the predicted hypotheses as pseudo-labels to attain better performance on all dev and test sets. The errors in pseudo-labels were further corrected with a language model using local prior matching (LPM) objective~\cite{hsu2020semi} and led to improvements in dev-other and test-other while the performance slightly degrades in dev-clean and test-clean. Our EAT model outperforms the self-training model on all evaluation sets as noted in table~\ref{tab:sota}.
\vspace{-0.2cm}
\begin{table}[h]
\caption{Comparison of SotA results in literature with EAT model}
\label{tab:sota}
\begin{tabular}{cccccc}
\hline 
\multirow{2}{*}{Method} & \multirow{2}{*}{Type} & \multicolumn{2}{c}{dev} & \multicolumn{2}{c}{test}\tabularnewline
 &  & clean & other & clean & other\tabularnewline
\hline 
\hline 
\multirow{2}{*}{Self-training} & Pseudo~\cite{Kahn_2020}& 5.41 & 20.31 & 5.79 & 21.63\tabularnewline
 & LPM~\cite{hsu2020semi} & 5.69 & 20.22 & 5.99 & 20.93\tabularnewline
\hline 
\multirow{2}{*}{Synthesis} & GST~\cite{rossenbach2020generating} & 7.4 & 25.7 & 7.9 & 26.7\tabularnewline
 & GCP~\cite{wang2020improving} & 4.1 & - & 4.1 & -\tabularnewline
\hline 
\multirow{2}{*}{Cycle} & ASR-TTS & 11.0 & 32.4 & 10.6 & 33.6\tabularnewline
 & EAT  & 4.3 & 14.9 & 4.3 & 15.2\tabularnewline
\hline 
\end{tabular}
\end{table}
GST~\cite{rossenbach2020generating} method focuses on attaining better synthesis quality by using GST speaker embeddings by training with TO data which is further used to train an ASR. This model attains 7.4\%WER and 7.9\%WER on dev-clean and test-clean which is relatively less due to the small language model used. In case of GCP, the authors synthesize speech and use consistency loss together to attain 4.1\%WER on dev-clean and 4.1\%WER on test-clean. This is understandably better than our EAT model since the GCP uses 460 hours of paired data while we use only 100 hours of paired data. The effect of penalizer, attention context scalar and other training strategies makes our EAT model attain 4.3\%WER on dev-clean and 4.3\% on test-clean. The model also attains the best performance on harder conditions such as dev-other and test-other.
\section{Conclusion}
In this work, we address the shortcomings of our previous ASR-TTS model and propose an enhanced ASR-TTS model. The proposed EAT model performs well on commonly used Librispeech task and shows its robustness to domain changes on BABEL-Pashto. The modification of SO with penalizer helped the model to improve on language related errors without hurting the acoustic information captured. The TO training by scaling attention context helped to improve on out-of-domain conditions such as Pashto and also brought gains in Librispeech. Training speech and text only (ST) together proved to be complementary and resulted in further performance improvement. The performance of EAT on Pashto is 51.6\%WER which is 2.7\% absolute less than with oracle 48.9\%WER. The model also attains 15.2\%WER on test-other which is 2.6\% absolute less than oracle performance. The model can be further enhanced on Librispeech by using 960 hours of unpaired speech and text data.

\bibliographystyle{ieeetr}
\bibliography{ref_new}

\end{document}